\documentclass[prl,twocolumn,superscriptaddress,floatfix,showpacs]{revtex4}
\usepackage{amsmath,amsfonts,epsf,dcolumn,natbib,graphicx}
\usepackage{ascii}
\usepackage{xcolor}
\usepackage{soul}

\newcommand{\be}{\begin{equation}}
\newcommand{\ee}{\end{equation}}
\newcommand{\beq}{\begin{equation}}
\newcommand{\eeq}{\end{equation}}
\newcommand{\bea}{\begin{eqnarray}}
\newcommand{\eea}{\end{eqnarray}}
\begin{document}
\bibliographystyle{plainnat}

\title{
Non-empirical Generalized Gradient Approximation Free Energy Functional for
Orbital-free Simulations
}

\author{Valentin V.~Karasiev}
\email{vkarasev@qtp.ufl.edu}
\author{Debajit Chakraborty}
\author{Olga A.~Shukruto}
\author{S.B.~Trickey}
\affiliation{Quantum Theory Project, 
Departments of Physics and of Chemistry, P.O. Box 118435, 
University of Florida, Gainesville FL 32611-8435}

\date{revised 08 October 2013}

\begin{abstract}
\noindent 
We report the first {\it wholly non-empirical} generalized gradient
approximation, non-interacting
free energy functional for orbital-free density functional theory  
and use that new functional to provide forces for finite-temperature 
molecular dynamics simulations in the warm dense matter (WDM) regime
The new functional provides good-to-excellent agreement with reference
Kohn-Sham calculations under WDM conditions at a minuscule fraction of 
the computational cost of corresponding orbital-based simulations. 
\end{abstract}


\maketitle
\renewcommand{\baselinestretch}{1.05}\rm

Compared to ordinary condensed matter, the warm dense matter (WDM)
regime \cite{Suhr2001,Driver.Militzer.2012} poses experimental
accessibility issues ({\it e.g.}\ inertial confinement fusion
hohlraums \cite{HEDLPreport.2009}) that make computational characterization of
WDM thermodynamics particularly significant.  Current practice, for example
Refs.\ \cite{Desjarlais02,Horner09},  
is {\it ab initio} molecular dynamics (AIMD) with Born-Oppenheimer 
electronic forces on the ions from finite-$T$  Kohn-Sham (KS) density
functional \cite{Mermin65,Stoitsov88,Dreizler89}
calculations.  Computational costs for KS-AIMD scale no better than $N_b^3$
per MD step, with $N_b$ the number of occupied KS
orbitals.  $N_b$ grows unfavorably with increasing $T$. 
KS-AIMD thus becomes prohibitively expensive 
at elevated $T$ and path integral Monte Carlo (PIMC)
simulations, which have comparable computational cost,  
come into play \cite{Driver.Militzer.2012}.

A long-standing potential alternative to KS-DFT, orbital-free DFT (OFDFT), 
would scale linearly with system size.  Use of OFDFT for WDM has been
limited by clearly inadequate functionals,
{\it e.g.}\ Thomas-Fermi \cite{Feynman..Teller.1949}, for the 
non-interacting kinetic energy (KE) part
${\mathcal T}_s$ of the free energy (though TF is, of course, 
the proper KS limit for high $T$ and
high material densities \cite{Horner09}).
Ground-state two-point orbital-free KE
functionals \cite{Xia.Carter.2012} 
are, unfortunately, of little utility for extension to WDM because
those two-point functionals which treat different material phases 
equally well are both parameterized and introduce substantial 
extra computational complexity.
Therefore we have focused on single-point functionals.

Here we provide a new, 
non-empirical, generalized gradient approximation (GGA)  ${\mathcal T}_s$ 
functional and its associated entropy functional.  They extend
and rationalize the constraint-based, mildly empirically parameterized 
GGA functionals recently published \cite{KST2}.  We show that 
the new functionals make OFDFT-AIMD competitive with finite-$T$ 
KS-AIMD calculations for accuracy and far faster.  
For deuterium in the WDM regime, the OFDFT AIMD and reference KS results 
agree well  at intermediate $T$, 
$ 6 \times 10^4 \rightarrow 1.8\times 10^5$ K.  
In the range $2 \times 10^5 \rightarrow 4 \times 10^6$ K, where computational
cost makes KS-AIMD data unavailable, the OFDFT AIMD and all-electron 
PIMC results \cite{Hu.Militzer..2011} compare well.  Similarly, 
the OFDFT-AIMD electron heat capacities for H at different material densities  
agree well with reference KS calculations up to 
$T=1\times 10^6$ K.

Ref.\ \cite{KST2} showed that well-behaved 
non-interacting free-energy GGA functionals should be 
defined in terms of distinct KE and   
entropic enhancement factors, $F_{\tau}(s_{\tau})$ and $F_{\sigma}(s_{\sigma})$, 
and showed that a useful approximation to their exact thermodynamic 
relationship is  
$F_{\sigma}(s_{\sigma}) \approx 2-F_{\tau}(s_{\sigma})$.
Each is a function of reduced density 
gradients with distinct explicit $T$-dependence, $s_{\tau}(n,\nabla n,t)$ and 
$s_{\sigma}(n,\nabla n, t)$, shown in detail in 
Ref.\ \cite{KST2}.  Here the reduced temperature is 
$t=T/T_{\mathrm F}=2/\beta [3\pi^2n({\mathbf r})]^{2/3}$, with 
$\beta=(k_{\mathrm B}{T})^{-1}$. Both $s_{\tau}$ and $s_{\sigma}$ go to 
the reduced density gradient  
familiar in exchange GGA functionals, 
$s(n,\nabla n)=|\nabla n|/\lbrace{2(3\pi^2)^{1/3}}n^{4/3}\rbrace$ as
$T \;\rightarrow \, 0$ K.
The GGA form for the non-interacting (KS system) free energy thus is 
\vspace*{-4pt} 
\be
\hspace*{-3.0pt}{\mathcal F}_{\mathrm s}^{\mathrm{GGA}}[n,T] \!=\! \int d{\mathbf r} %
\tau_0^{\mathrm{TF}}(n)\lbrace \xi(t) F_{\tau}(s_{\tau}) %
-\zeta(t) F_{\sigma}(s_{\sigma}) \rbrace \, ,
\label{E1}
\ee
%
%
where $\tau_0^{\mathrm {TF}}$ is the zero-$T$ TF KE density.  The functions  
$\xi(t)$ and $\zeta(t)$ are smooth, well-behaved combinations of Fermi-Dirac
integrals, with forms given explicitly in \cite{KST2}.  The unaddressed
problem in Ref.\ \cite{KST2}, which we resolve here, is how to get a reliable, 
wholly non-empirical representation of $ F_{\tau}$.

In Eq.\ (\ref{E1}), $t$ appears such that 
the $T = 0$ K limit of the GGA free-energy is a 
ground-state OF-KE functional  
defined by the enhancement factor $F_{\tau}(s)$, that is \vspace*{-8pt} 
\be
\lim_{ T \rightarrow 0} {\mathcal F}_{\mathrm s}^{\mathrm {GGA}}[n,T]=
\int d{\mathbf r}
\tau_0^{\mathrm{TF}}(n) F_{\tau}(s)={\mathrm T}_{\mathrm s}^{\mathrm{GGA}}[n]\,.
\label{E2}
\ee
Therefore the enhancement factor $F_{\tau}(s)$ and the functional
Eq.\ (\ref{E2}) are subject to 
$T=0$ K KE constraints.  These include 
(i) recovery of the second-order 
gradient expansion (GE) in the small-$s$ limit \cite{Hodges73}, 
$F_{\tau}(s)\approx 1+(5/27)s^2$; 
(ii) a non-negative Pauli potential 
\cite{LevyOu-Yang88,LevyPerdewSahni84,Herring86}, 
\be
v_{\theta}([n];{\mathbf r}):=\frac{\delta {\mathrm T}_\theta}{\delta n} \equiv 
\frac{\delta ({\mathrm T}_{\mathrm s}[n]-{\mathrm T}_{\mathrm{vW}}[n])}{\delta n({\mathbf r})} \geq 0\,, 
\;
 \forall \; {\mathbf r} \, ,
\label{E3}
\ee
with 
${\mathrm T}_{\mathrm {vW}}[n]=\int d{\mathbf r}\tau_0^{\mathrm {TF}}(n) (5s^2/3)$
the von Weizs\"acker (vW) functional  \cite{Weizsacker}; 
and (iii) recovery of vW behavior in the large-$s$ limit. 

Constraint (i) guarantees a correct description for 
uniform and slow-varying densities. 
As shown in Refs.\ \cite{PRB80,CPC-VVK-SBT-2012}, positivity of 
$v_\theta$ is required to achieve  
molecular and solid binding. Constraint (iii)  
follows from the character of charge densities far
from any nucleus and the so-called IP theorem \cite{LevyPerdewSahni84}.  
However, the 
analytical form of the KE enhancement factor is a matter of design 
choice, sometimes motivated by the conjointness conjecture \cite{LeeLeeParr91},
to wit $F_{\tau}(s) \propto F_{\mathrm x}(s)$. Thus, the 
non-empirical APBEK \cite{CFDS11} 
$T$ = 0 K functional uses the PBE X enhancement factor 
form \cite{PerdewBurkeErnzerhof96}.
Manifestly it violates constraint (iii).  
As to (i), the GE coefficient for APBEK is 0.23889, which corresponds to the 
modified gradient expansion \cite{CFDS11}.  But  $v_\theta$ from 
APBEK violates constraint (ii) in that $v_\theta^{\mathrm{APBEK}}$ has 
negative singularities at nuclear positions. 
The behavior of $v_{\theta}$ near a 
nucleus, $r\approx 0$, follows from the Kato nuclear-cusp 
condition \cite{Kato57}
\be
n(r)\sim e^{-2Zr}=(1-2Zr)+O(r^2)\,.
\label{E4}
\ee
Thus $v_{\theta}^{\mathrm {APBEK}}(r)\sim a/r$ with $a<0$ for $r\approx 0$. 

To satisfy constraints
(i) and (ii) simultaneously and  incorporate (iii) therefore requires a more 
flexible form.  Constraint (iii) also occurs in the 
VT$\{$84$\}$ X enhancement factor \cite{VT84}, so we adopt a suitably
modified form for $F_{\tau}$, 
\be
F_{\tau}^{\mathrm {VT84F}}(s)=1-\frac{\mu s^2e^{-\alpha s^2}}{1+\mu s^2} + (1-e^{-\alpha s^{m/2}})
(s^{-n/2}-1)+\frac{5s^2}{3}
\,,
\label{E5}
\ee
with $m=8$, $n=4$. (``F'' in ``VT84F'' denotes this free-energy adaptation.) 
The last term in Eq.\ (\ref{E5}) provides the correct large-$s$
limit, constraint (iii). The parameters $\mu$ and $\alpha$ then 
must be determined from constraints (i) and (ii).
Expansion of Eq. (\ref{E5}) at small-$s$ gives 
$F_{\tau}^{\mathrm {VT84F}}(s)=1+(5/3+\alpha-\mu)s^2+O(s^4)$.
Constraint (i) imposes a relation between the two parameters,
 $\alpha=\mu-5/3+5/27$. Evaluation of the Pauli potential 
for small-$r$ from the density 
Eq.\ (\ref{E4}), shows that the singular term $a/r$ becomes 
marginally positive for $\mu=2.778$. That gives $\alpha=1.2965$. 
Eq.\ (\ref{E5}) then 
fixes the kinetic and entropic enhancement factors 
in the free-energy functional Eq.\ (\ref{E1}), 
$F_{\tau}^{\mathrm{VT84F}}(s_{\tau})$ 
and  
$F_{\sigma}^{\mathrm {VT84F}}(s_{\sigma})=2-F_{\tau}^{\mathrm {VT84F}}(s_{\sigma})$.
For comparison, we also built the non-interacting free-energy functional 
APBEF from the zero-$T$ APBEK KE \cite{CFDS11} by use of the same prescription, 
that is 
$F_{\tau}^{\mathrm{APBEF}}(s_{\tau})=1+\mu s_{\tau}^2/(1+s_{\tau}^2\mu/\kappa)$
and $F_{\sigma}^{\mathrm {APBEF}}(s_{\sigma})=2-F_{\tau}^{\mathrm {APBEF}}(s_{\sigma})$ 
with $\mu=0.23889$, $\kappa=0.804$.

\begin{figure}
\includegraphics*[width=6.5cm]{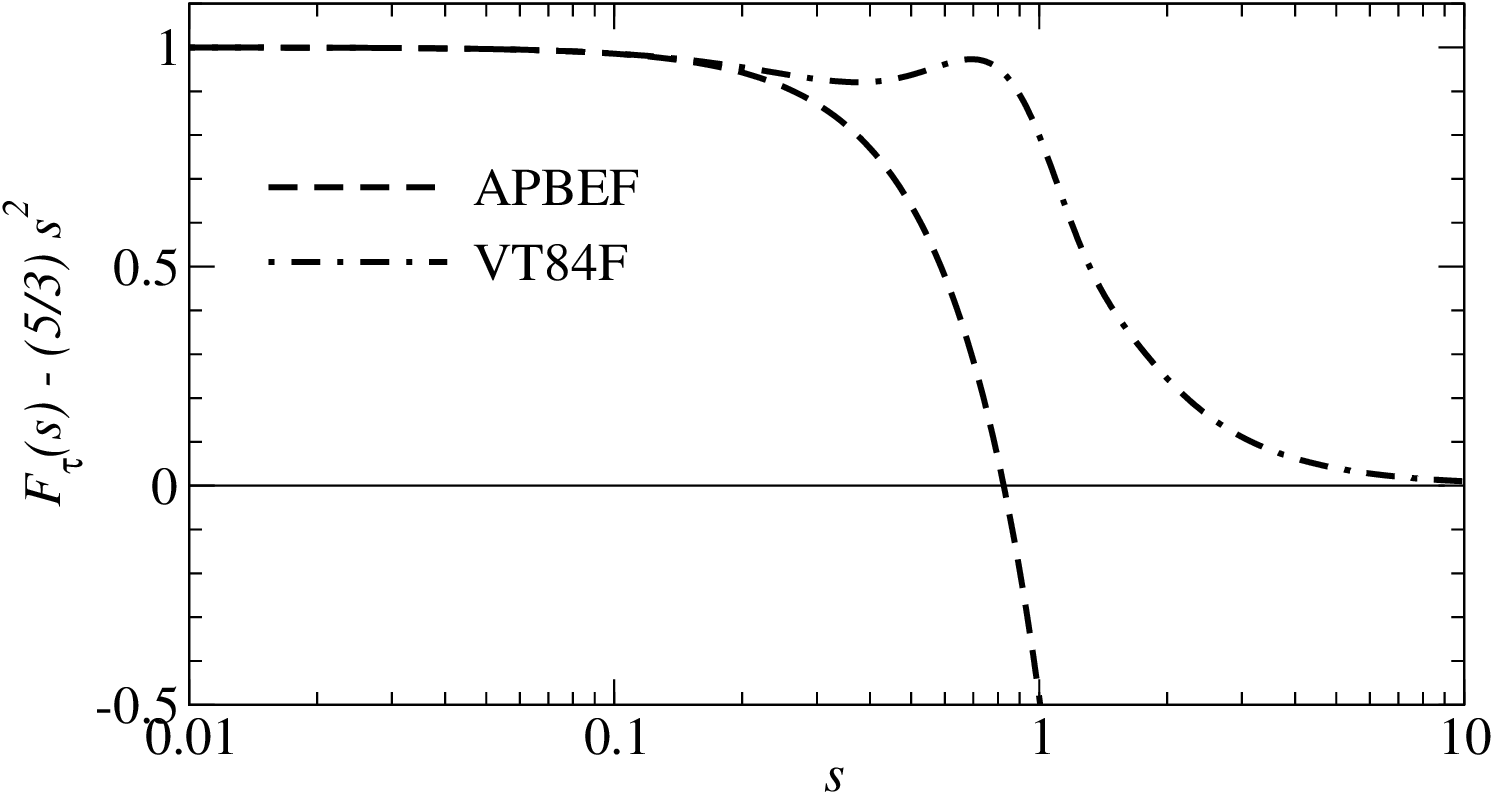}
\caption{
VT84F and APBEF Pauli term enhancement factors $F_{\tau}(s)-(5s^2/3)$ 
as a function of $s$ ($T$ = 0 K).
}
\label{Ftheta}
\end{figure}

Fig.\ \ref{Ftheta} shows the two main differences between the VT84F and 
APBEF  Pauli enhancement factors,  
$F_{\tau}(s)-(5s^2/3)$.  For VT84F, $F_{\tau}(s)-(5s^2/3)$
is non-negative and vanishes at large-$s$ and has positive slope 
near $s\approx 0.39$ to provide the correct sign of the corresponding
$v_\theta^{\mathrm{VT84F}}$ near nuclear sites. APBEF has neither feature.  
At small-$s$, both functions have similar behavior 
defined by the gradient expansion with similar coefficients.

We have implemented these functionals in a modified version of the
{\sc Profess} \cite{Hung..Carter10} code which we have 
interfaced to the {\sc Quantum Espresso}
code \cite{QEspresso} to support KS and OFDFT AIMD calculations on the same
footing \cite{ProfessAtQE}.   The 
data in Table \ref{tab:table1} illustrate the critical importance of 
satisfying  constraint Eq. (\ref{E3}). (Both these calculations used 
Perdew-Zunger local density approximation (LDA) exchange-correlation (XC) \cite{PZ81}.)  At $T =0$ K, the VT84F KE functional  
gives binding in sc-H and fcc-Al with lattice constants 
underestimated by about 6\% 
for sc-H and about 2\% for fcc-Al. 
The APBEK functional has typical ordinary GGA KE functional behavior. It fails 
to yield binding because of violation of 
constraint Eq. (\ref{E3}) \cite{PRB80}.  The bulk moduli from VT84F, however, 
are higher than the reference KS values.

\begin{table}
\vspace*{-8pt}
\caption{\label{tab:table1}
Equilibrium lattice constants $a$ and 
bulk moduli $B$ 
calculated with the 
VT84F KE functional. OFDFT calculations with 
APBEK do not yield equilibrium configurations.
KS LDA values are shown for comparison.
All OFDFT calculations with $T$-independent LDA XC \cite{PZ81}.
}
\begin{ruledtabular}
\begin{tabular}{lcc}
System/Method &   $a$ (\AA)  & $B$ (GPa) \\
\hline
\underline{sc-H} &&\\
OFDFT (VT84F+LDA)   & 1.353 & 175.3  \\
KS (LDA)\cite{KST2}&  1.446 & 108.4  \\
\hline
\underline{fcc-Al} &&\\
OFDFT (VT84F+LDA)   & 4.095 & 120.4  \\
KS (LDA)\cite{Boettger.Trickey.1996}
                   &  4.020 & 79.66  \\

\end{tabular}
\end{ruledtabular}
\end{table}

\begin{figure}
\includegraphics*[width=6.5cm]{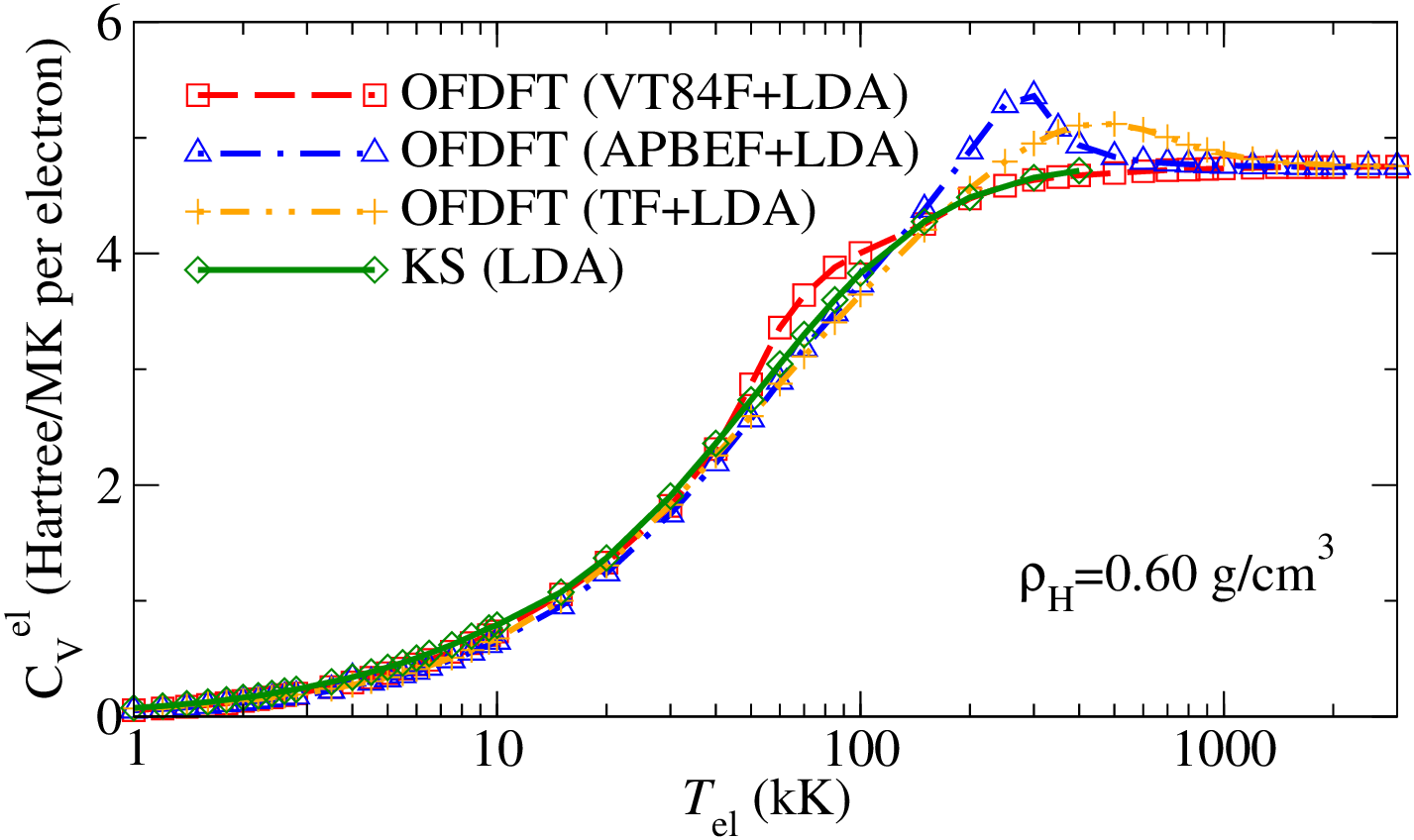}
\includegraphics*[width=6.5cm]{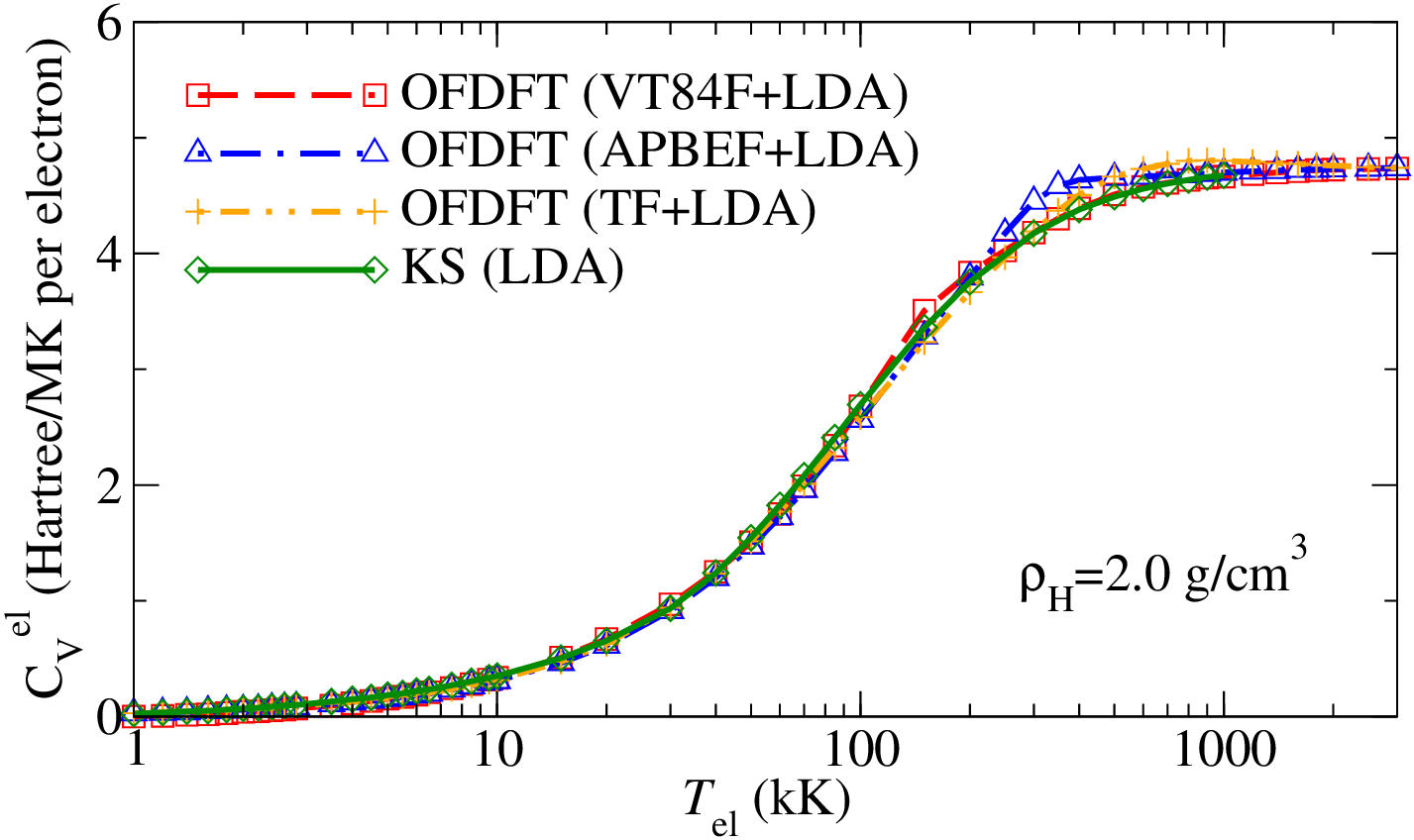}
\caption{
Electronic heat capacity, $C_{\mathrm V}^{\mathrm{el}}$, as a 
function of electronic $T$ for sc-H at material density 
$\rho_{\mathrm H}$=0.60 and 2.0 g/cm$^3$.
}
\label{C_v-vs-T.HR1.0}
\end{figure}

To test the OF functionals at finite $T$, we started from
static calculations with cold nuclei and hot electrons. Such a
situation arises, for example, when a target is irradiated by a 
femtosecond laser pulse \cite{Ernstorfer.Science.2009}.   Calculations were
done for sc-H at material density $\rho_{\mathrm H}=$ 0.60 and 2.0
g/cm$^3$ ($r_s=$1.650 and 1.105 bohr respectively) with 64 atoms in the 
simulation cell. The reference KS calculations used 8 atoms
in a supercell and a $13\times 13\times 13$ Monkhorst-Pack Brillouin zone grid
\cite{MonkhorstPack76}.  Our transferable PAW data set \cite{PRE86} was
employed in the KS calculations, and a similarly transferable local
pseudopotential \cite{KST2} was used in the 
OFDFT calculations. For this stage of testing, ordinary PZ LDA XC 
again was used \cite{PZ81}.  
Owing to machine-time limitations, we were able to complete KS 
calculations only up to $T = 4\times 10^5$ K for 
$\rho_{\mathrm H} = $0.60 g/cm$^3$ and 
to 10$^6$ K for $\rho_{\mathrm H} = $2.0 g/cm$^3$. 

Fig.\ \ref{C_v-vs-T.HR1.0}
compares the electronic heat capacity, $C_{\mathrm V}^{\mathrm{el}}=(\partial
E^{\mathrm{el}}/\partial T_{\mathrm{el}})_{\mathrm V}$, where $E^{\mathrm{el}}$ is 
the electronic  internal energy and $T_{\mathrm{el}}$ is the electronic 
temperature and the units are per atom.  
At low $T$,  $C_{\mathrm V}^{\mathrm{el}}$ goes linearly with $T$. 
In the high-$T$ limit, it goes to the classical
ideal gas value, $(3/2)k_{\mathrm B}=4.750$ Hartree/megaK per particle. 
Values from the new 
VT84F functional agree quite well with the KS
data for the whole range of $T$, except for a small deviation near
80 kK for $\rho_{\mathrm H}=0.60$ g/cm$^3$.  Both the VT84F and KS values
exhibit only a weak dependence on material density and   
converge slowly to the TF limit, which is reached  at $T \approx 1500$ kK. 
By comparison,  $C_{\mathrm V}^{\mathrm{el}}$ values from the APBEF and TF
functionals agree well with the KS data for low $T$,  up to
about 60 kK for $\rho_{\mathrm H} = $2.0 g/cm$^3$.   
But for the lower density,  
the APBEF results deviate from the KS data up to 20\%
for $T_{\mathrm{el}}$ between approximately
150 kK and 600 kK, whereas the TF results have a comparable deviation
in the range of about 200 kK $\le T \le$  900 kK.
A technical point is that the second derivative discontinuity of fits 
used in the OFDFT calculations (see Ref.\ \cite{KST2}) affects the 
OFDFT results for $C_{\mathrm{V}}^{\mathrm{el}}$ at $T \approx T_{\mathrm F}/2$.  

The second finite-$T$ test of our new functional was to calculate 
the deuterium equation of state (EOS) in the WDM regime 
\cite{McMahonMoralesPierleoniCeperley}.  
All the AIMD simulations were performed with 64-512 atoms in the 
simulation cell (depending on material density) using the 
$NVT$ ensemble regulated by the Andersen thermostat.
For KS calculations at $T \le 31,250$ K, we used
a $3\times 3\times 3$ Monkhorst-Pack k-grid \cite{MonkhorstPack76},
while for higher $T$ a single $\Gamma$-point was used.
All the calculations used an  
explicitly $T$-dependent LDA (TLDA) XC functional
\cite{PDW2000}; see Ref.\ \cite{PRE86} for justification. 

\begin{figure}
\includegraphics*[width=6.5cm]{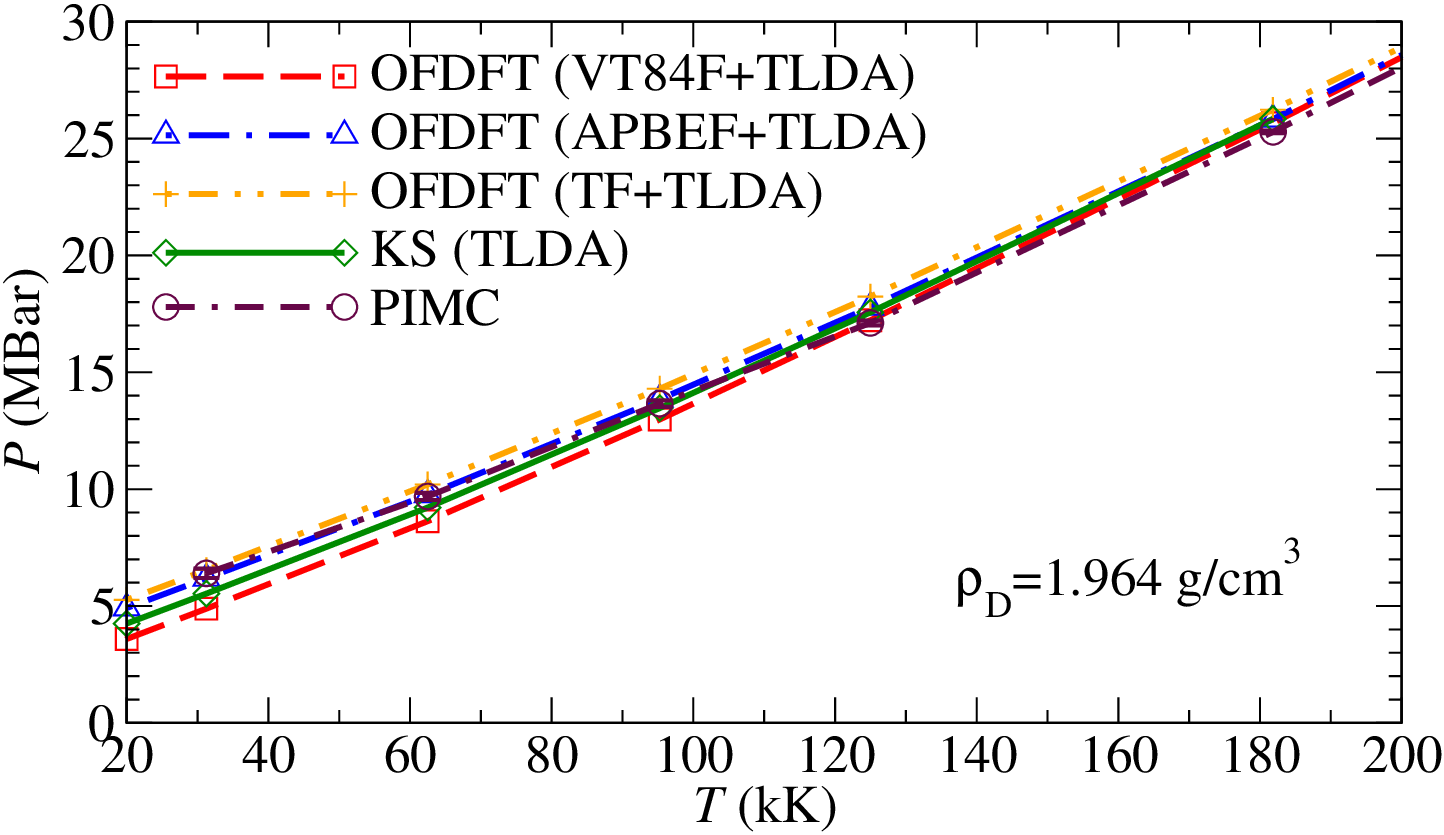}
\includegraphics*[width=6.5cm]{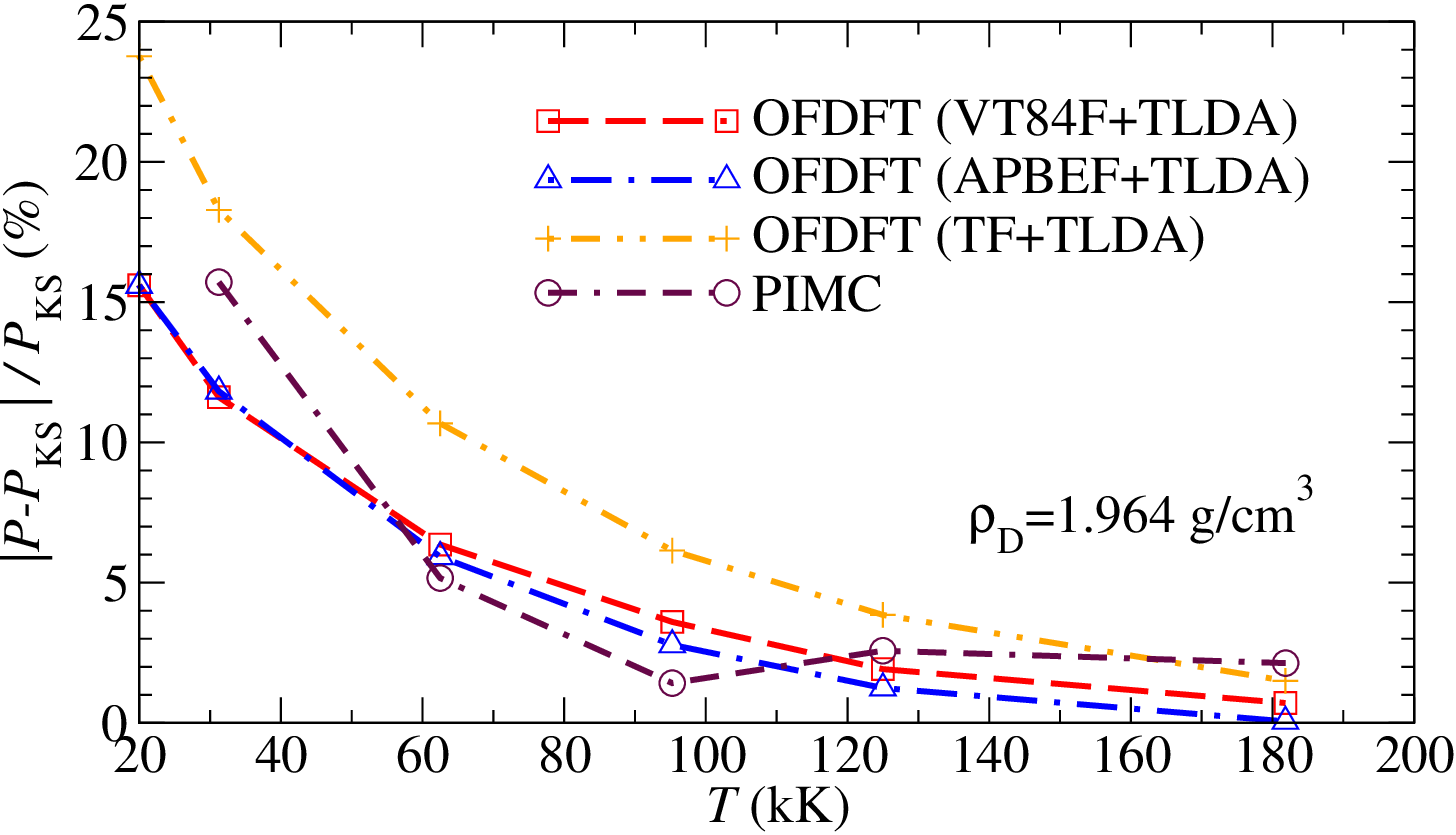}
\caption{
Upper panel: pressures for OFDFT and KS AIMD, both  with  
explicitly $T$-dependent XC \cite{PDW2000}) compared with 
PIMC \cite{Hu.Militzer..2011} results 
for Deuterium at 
$\rho_{\mathrm D}=1.964$ g/cm$^3$ ($r_{\mathrm s}=1.40$ bohr).
Lower panel: relative differences of OFDFT and PIMC pressures with respect
to KS values.
}
\label{P-vs-T.D128R1.964}
\vspace*{-8pt}
\end{figure}

The upper panel of Fig.\ \ref{P-vs-T.D128R1.964} compares pressures 
for deuterium at $\rho_{\mathrm D} =$1.964 g/cm$^3$ ($r_{\mathrm s}=1.40$ bohr) 
from OFDFT and KS AIMD simulations,  along with PIMC results.  
Our VT84F functional tends to underestimate
the pressure while both 
TF and APBEF overestimate it. 
However, our new functional reduces the error at $T=200$ kK to 15\% compared
to the TF error of 24\%.  Note that APBEF, which fails to predict an 
equilibrium ground state, nevertheless gives about the same relative 
pressure error as VT84F, hence provides an inconsistent description.   
The error in the OFDFT values decreases with 
increasing $T$, such that  
at $T=95,250$ K that error is about 3 \% for the two GGAs  
versus  6 \% for TF.  At $T=181,825$ K (the highest $T$ for which we were 
able to complete the KS AIMD simulation), that error is 1.5 \% for TF 
compared to tenths of a percent for VT84F (and for APBEF
as well). Comparison of PIMC to KS gives relative differences of 
essentially the same magnitude as the OFDFT calculations which use
the new functionals.  At  the lowest temperature, $T=31,250$ K, 
PIMC overestimates the pressure by 15\%, with the error decreasing
rapidly with increasing $T$.

\begin{figure}
\includegraphics*[width=6.5cm]{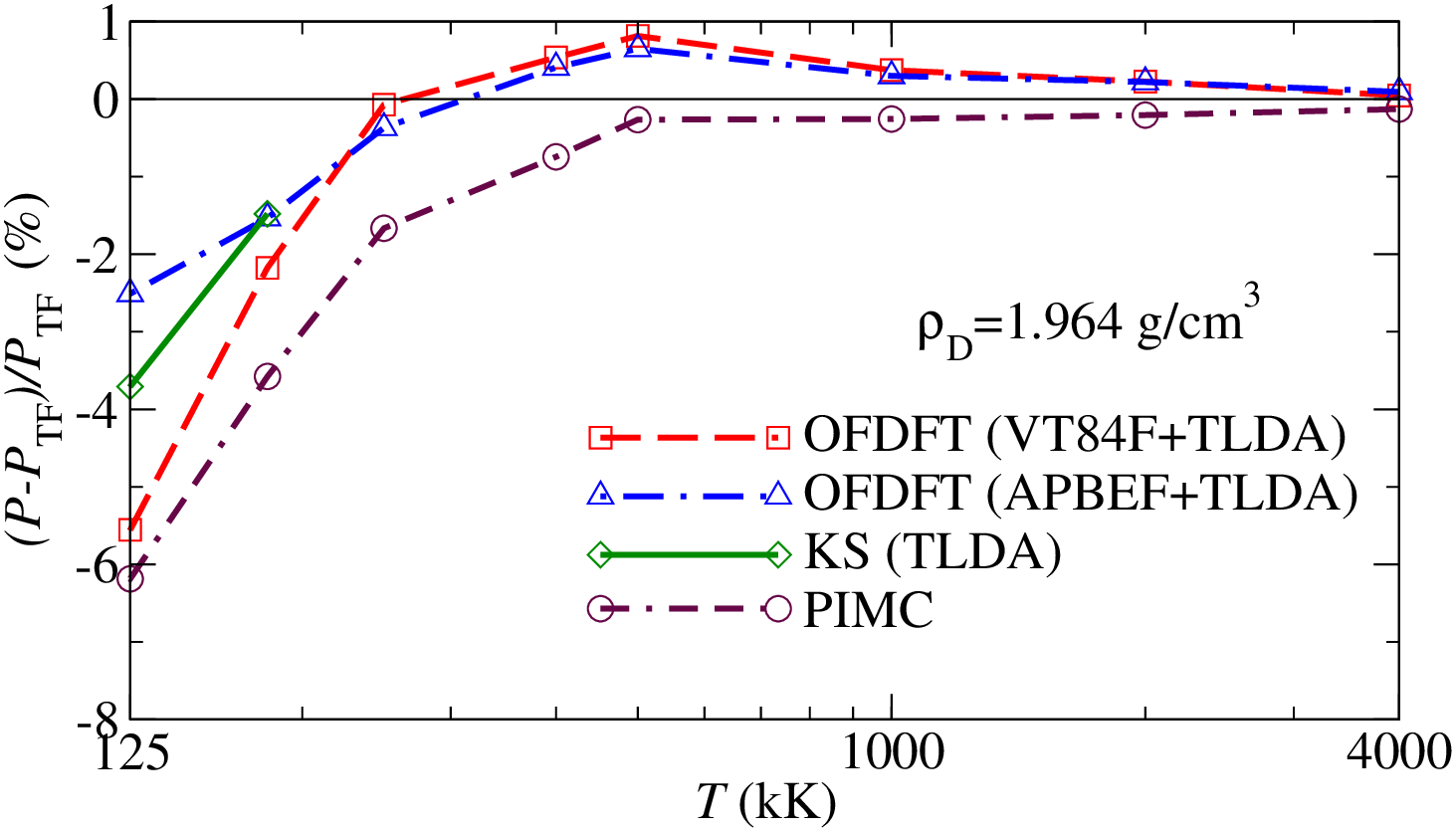}
\caption{
Excess pressure relative to the TF model for  OFDFT (APBEK and VT84F 
functionals), KS and PIMC \cite{Hu.Militzer..2011},
for deuterium at material density $\rho_{\mathrm D}=1.964$ g/cm$^3$ 
($r_{\mathrm s}=1.40$ bohr). 
}
\label{excess-P-rel-TTF}
\end{figure}

In the high-$T$ TF limit, the system goes over to a fully 
ionized electron-ion plasma.
Fig.\ \ref{excess-P-rel-TTF} shows the excess pressure relative to the TF model
for $125,000 \le T \le 4,000,000$ K.
For $T=125,000$ and $181,825$ K, where KS data are available, 
both VT84F and APBEF,
provide excellent agreement (within about 2\%). Our OFDFT results also are
in reasonably good  
agreement with the PIMC data (almost within the margin of numerical error).

\begin{figure}
\includegraphics*[width=6.5cm]{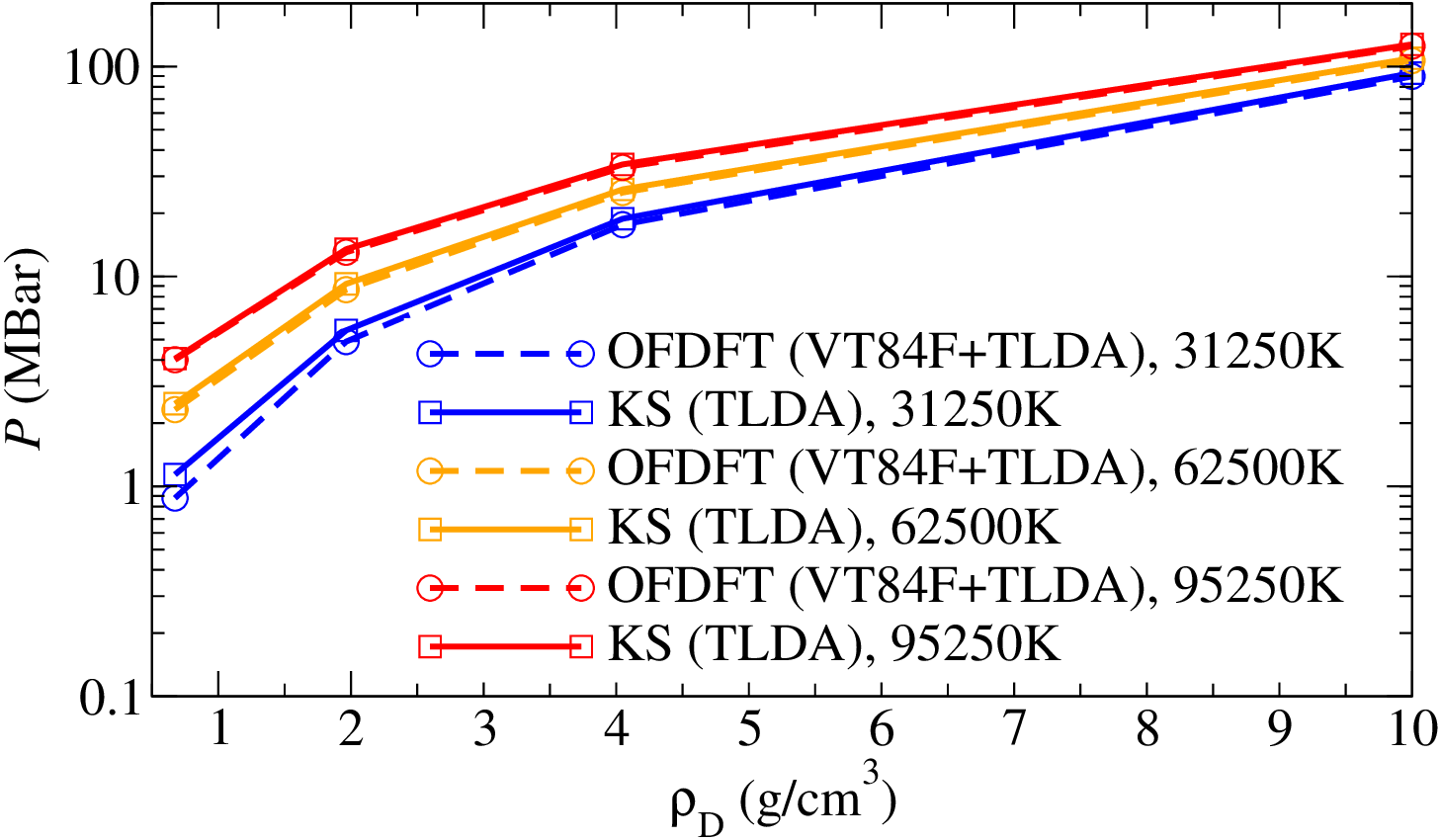}
\caption{
Pressure vs.\ material density for selected 
temperatures calculated by OFDFT and KS AIMD for deuterium.
}
\label{P-vs-Rho_D}
\end{figure}

Fig.\ \ref{P-vs-Rho_D} compares KS and OFDFT
pressures for deuterium as a function of $\rho_{\mathrm D}$ for 
three temperatures. The
small deviations of the values from the VT84F functional with respect
to the KS values at lowest density, 
$\rho_{\mathrm D}=0.674$ g/cm$^3$, T=$31,250$ K, diminish quickly 
with increasing $\rho_{\mathrm D}$ or increasing $T$.

\begin{figure}
\includegraphics*[width=6.5cm]{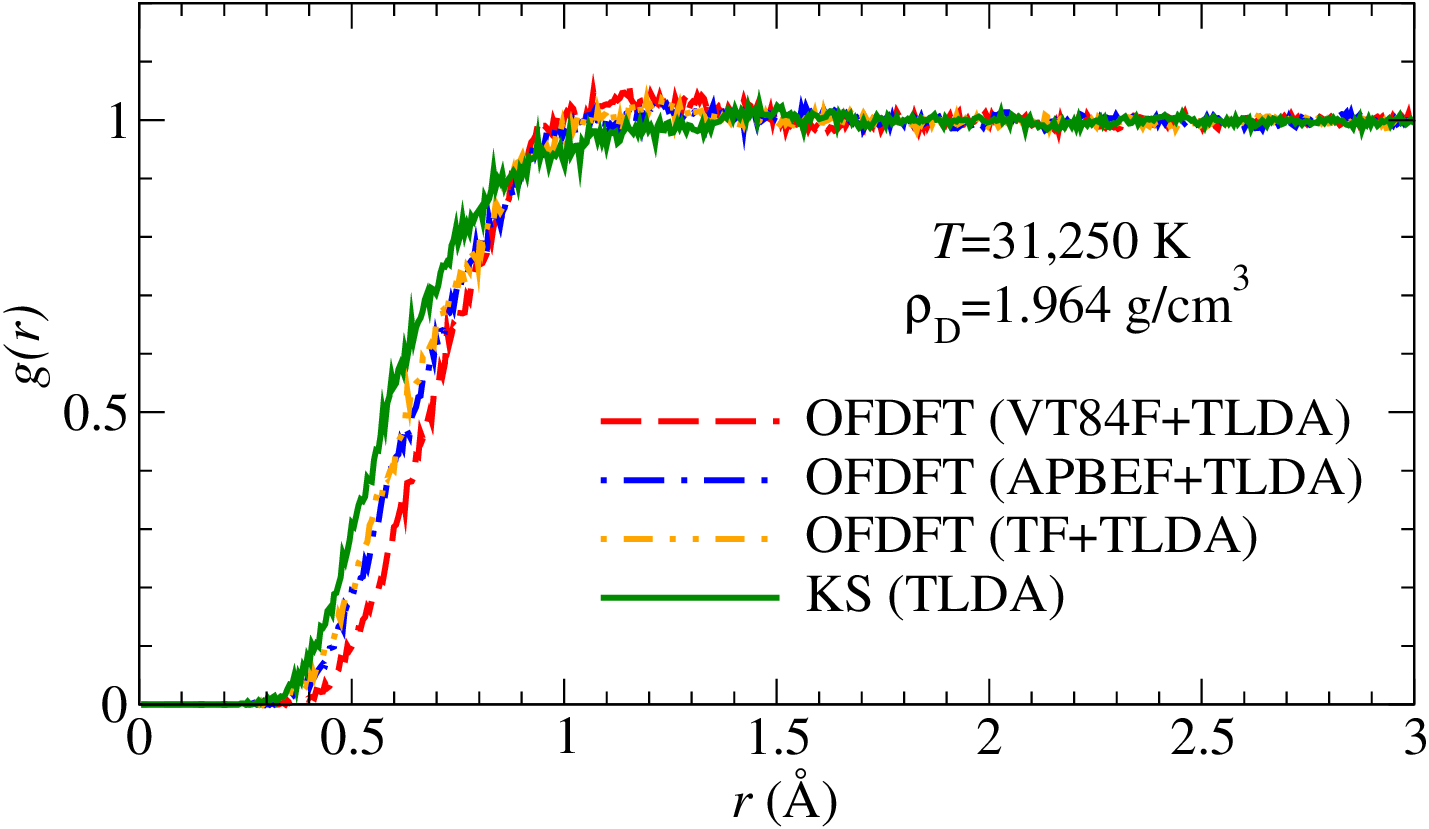}
\includegraphics*[width=6.5cm]{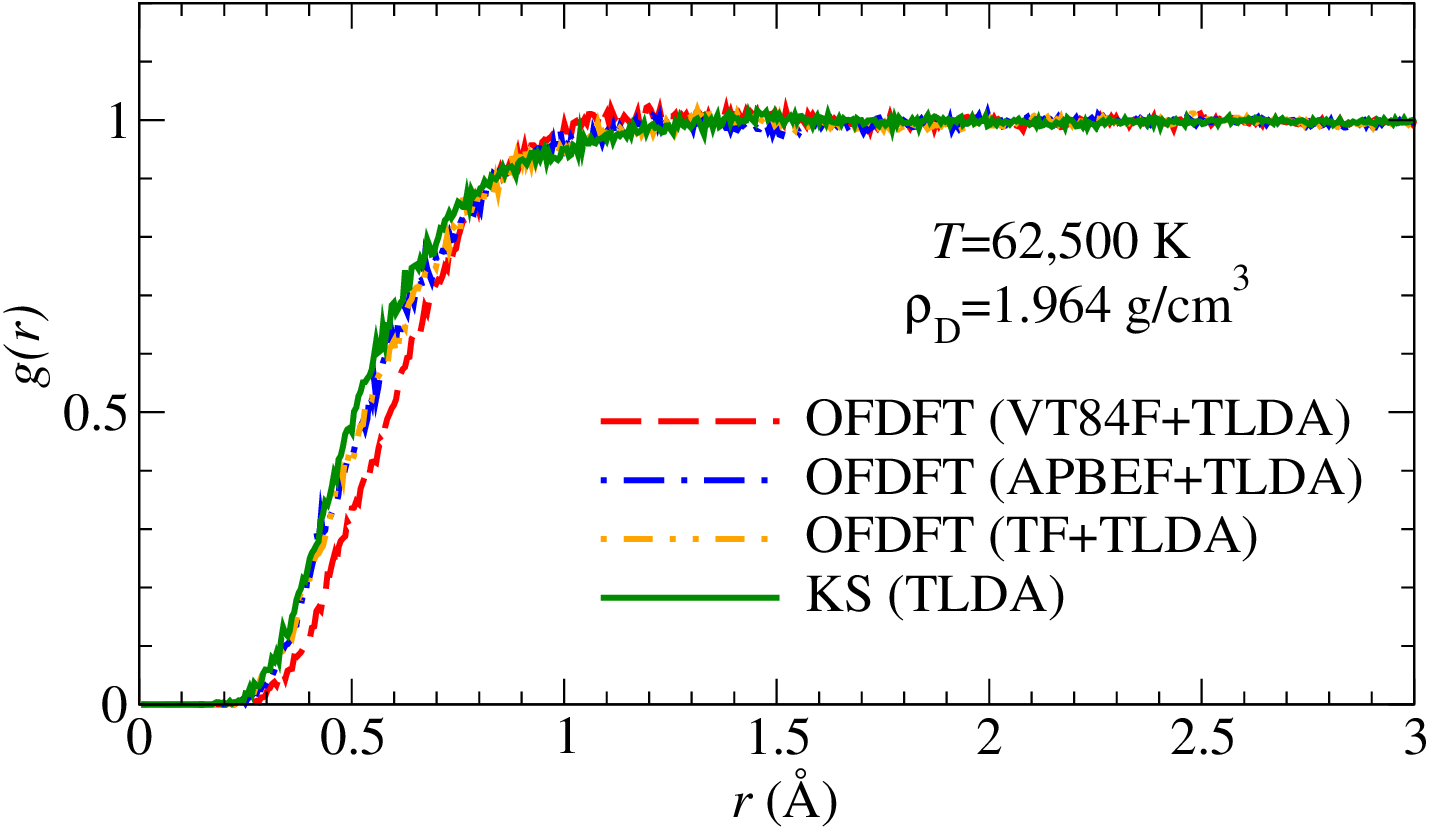}
\caption{
The OFDFT and KS ion pair-correlation function for $T=31,250$ K 
(upper panel) and $T=62,500$ K (lower panel. 
}
\label{g-of-r.D128R1.964}
\end{figure}

Fig.\ \ref{g-of-r.D128R1.964} compares KS and OFDFT ion pair-correlation
functions (PCF) for two temperatures. The upper panel ($T=31,250$ K) 
demonstrates that 
all the OFDFT calculations predict structural properties at this temperature 
in reasonable 
agreement with the KS results, except for some discrepancies 
(peaks) near $ r = 1.0 $\AA.
We suspect, but have not been able to confirm, that those peaks are 
related to nuclear site singularities in the GGA Pauli potential, Eq.\ 
(\ref{E3}).  Those singularities could lead to peaks such as seen in
 hard- or soft-sphere liquid PCFs \cite{AndersonWeeksChandler71}.
Note also that the peaks are consistent with the overly large bulk
moduli via the compressibility sum rule \cite{Rowlinson65}. 
In any event, for $T=62,500$ K and above, the agreement 
between OFDFT and KS PCFs becomes satisfactory.

\begin{figure}
\includegraphics*[width=6.5cm]{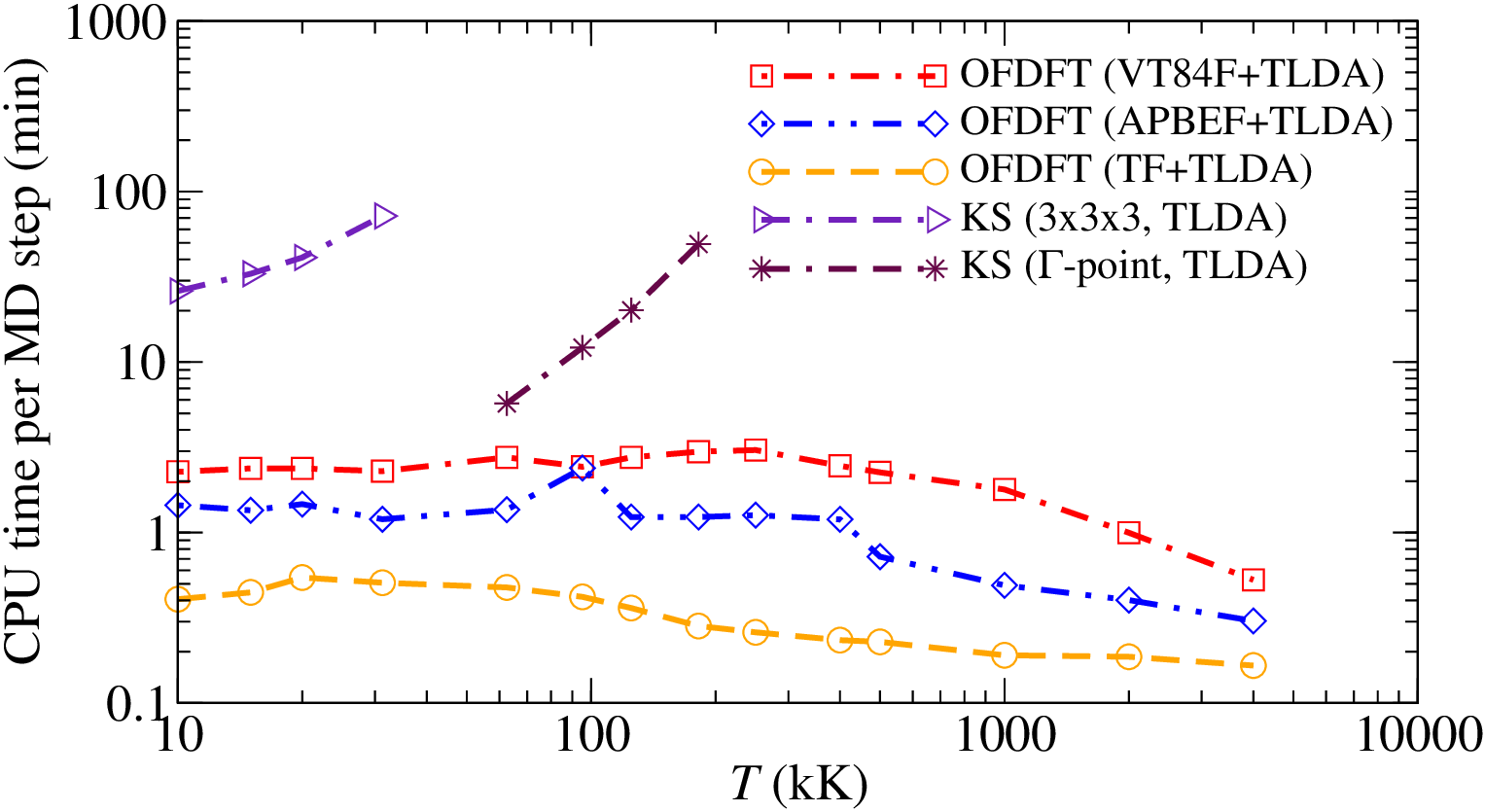}
\caption{
CPU time per AIMD step as a function of $T$ for OFDFT-MD calculations
compared to the KS-MD data. Deuterium at 
$\rho_{\mathrm D}=1.964$ g/cm$^3$ ($r_{\mathrm s}=1.40$ bohr), 
128 atoms in simulation cell.
}
\label{cputime-vs-T.D128R1.964}
\end{figure}

Comparison of computational times per AIMD step for OFDFT and KS is in 
Fig.\ \ref{cputime-vs-T.D128R1.964}. The calculations 
were done on a single CPU to provide the most favorable case for KS (no
parallel overhead).  The OFDFT timings  
are essentially independent of $T$ and faster than corresponding KS AIMD runs
by from one to two orders of magnitude for the range of $T$ shown. 
In practice, the KS calculations typically need 8 to 64 CPUs for
reasonable turn-around.  In that case, the OFDFT advantage is substantially
greater.  

In summary, we have presented a new, wholly non-empirical parameterization
of a ground-state orbital-free KE functional and used it to generate 
new kinetic and entropic non-interacting free-energy functionals.  These
new functionals have several virtues.  First, the ground
state part gives a reasonable description
of the ground-state solid for sc H and fcc Al, something not achieved
by any other non-empirical KE GGA.  Second, the consequent
free-energy functionals give good WDM
properties for sc-H in the static lattice case ({\it e.g.} electronic
heat capacity) and  provide a competitive-quality AIMD simulation
of the deuterium EOS. All of this is with the long-promised computational speed
advantage of OFDFT.  

We thank T.\ Sjostrom for valuable discussions.  
This work was supported by the U.S.\ Dept.\ of Energy TMS grant DE-SC0002139.
We also acknowledge the University of Florida High-Performance Computing 
Center for providing computational resources and technical assistance.

\end{document}